\documentclass[]{IEEEtran}
\IEEEoverridecommandlockouts 
\ifCLASSINFOpdf
\usepackage[pdftex]{graphicx}
\graphicspath{{./Images/}}
\DeclareGraphicsExtensions{.pdf,.jpeg,.png}
\else
\fi

\usepackage[cmex10]{amsmath}
\usepackage{siunitx}
\usepackage{cite}
\usepackage{psfrag}
\usepackage[utf8]{inputenc}
\usepackage[T1]{fontenc}
\usepackage{amsmath,amsfonts,amsbsy,amssymb}
\usepackage{mathabx}
\usepackage{mathrsfs}
\usepackage[nolist]{acronym}
\usepackage{tabularx}
\usepackage{amssymb}
\usepackage{amsmath}
\usepackage{graphicx}
\usepackage{cite}
\usepackage{multirow}
\usepackage{wasysym}
\usepackage{multirow}
\usepackage{float}
\usepackage{color}
\usepackage{subcaption}

\newtheorem{proposition}{Proposition}
\newtheorem{corollary}{Corollary}

\begin{document}
\title{Storage Management in \\ Modern Electricity Power Grids}

\author{Pedro H. J. Nardelli and Hirley Alves
\thanks{The authors are with the Centre for Wireless Communications (CWC) at University of Oulu, Finland. This work is partly funded by Finnish Academy (n. 271150 and n. 303532) and CNPq/Brazil (n.490235/2012-3) as part of the joint project SUSTAIN and by Strategic Research Council/Aka BCDC Energy (n.292854). Contact: pedro.nardelli@oulu.fi}%
}

\maketitle

\begin{abstract}
This letter introduces a method to manage energy storage in electricity grids.
Starting from the stochastic characterization of electricity generation and demand, we propose an equation that relates the storage level for every time-step as a function of its previous state and the realized surplus/deficit of electricity.
Therefrom, we can obtain the probability that, in the next time-step: (i) there is a generation surplus that cannot be stored, or (ii) there is a demand need that cannot be supplied by the available storage.
We expect this simple procedure can be used as the basis of electricity self-management algorithms in micro-level (e.g. individual households) or in meso-level (e.g. groups of houses).
\end{abstract}

\vspace{1ex}


\section{Introduction}
The transition from centralized electricity power grid toward a more decentralized one is posing different challenges \cite{nardelli2014models}.
In specific terms, the intermittence of low-carbon renewable sources like wind and solar poses an interesting management task \cite{Barton2004}: periods of high demand might not match with supply.
Energy storage then becomes an important capability for managing the possible mismatches.

In this letter, we use the stochastic characterization of both electricity generation and electricity demand in different periods during the day  (e.g. hour-by-hour).
We follow the literature (e.g. \cite{Celik2004,Carpaneto2008,Munkhammar2015,Shariatkhah2016}) and assume that both may be modeled as independent random variables at each period.
Depending on surplus/deficit outcome, the storage managing entity shall estimate the situation for next period.

Our goal here is to understand the conditions that a system composed by a intermittent source plus a storage device can be self-managed.
We propose an equation to evaluate the probability that the system is not self-sufficient either by generating more electricity than can be stored, or by consuming more than it is supplied.
We numerically demonstrate this procedure using two examples: a virtual utility that aggregates several households with a hydro power plant as storage and wind as generation, and a household with battery and solar panels.

We interpret these results as a promising way to self-manage electricity looking at its use value.
They also indicates promising ways to implement cooperative sharing of surpluses, mainly when weather forecast is available.

\section{Theoretical framework}
Consider a scenario composed by: an entity that demands electricity, an entity that generates electricity, and an entity that stores energy to balance possible surpluses/deficits.
We assume a discrete time setting where each time step $t \in \mathbb{N}$.
From previous studies \cite{nardelli2014models,Barton2004,Carpaneto2008,Celik2004,Shariatkhah2016,Munkhammar2015}, we model to each time step as follows.
\begin{itemize}
	\item Demand: $D(t)\geq 0$ is a random variable;
	%
	\item Generation: $G(t)\geq 0$ is a random variable;
	%
	\item Storage: $S_\textup{min} \leq S(t)\leq S_\textup{max}$ is a random variable limited by its minimum $S_\textup{min} \geq 0$ and maximum $S_\textup{max} > S_\textup{min}$.
\end{itemize}

\begin{proposition}
\label{prop_storage_equation}
	The storage state at time step $t$ is given by:
	\begin{equation}
		\label{eq-storage}
		S(t) = \min\left[S_\textup{max}, \max\left[S(t-1) + B(t) \right], S_\textup{min}\right],
	\end{equation}
	where $B(t) = G(t) - D(t)$ is a random variable indicating the balance between supply and demand during time step $t$; note that $B(t) \in \mathbb{R}$.
\end{proposition}

\begin{IEEEproof}
The storage situation at time step $t$ is given by the realization of supply and demand at $t$, considering the storage situation in previous time step $t-1$.
Then: $S(t) = S(t-1) + B(t)$.
If the storage upper and lower limits $S_\textup{max}$ and $S_\textup{min}$ are considered, we find \eqref{eq-storage}.
\end{IEEEproof}

Note that the storage situation $S(t)$ comes after the realization of generation $G(t)$ and demand $D(t)$.
These latter variables are the inputs of this model and their actual distribution depends on the season, the hour of the day, whether is a working day etc.
Equation \eqref{eq-storage} works regardless since it is based on the realized values.

If the goal is prediction, the storage managing entity can estimate the upcoming conditions $S(t+1)$ if it knows (i) the own state $S(t)$ and (ii) the distributions of $G(t)$ and $D(t)$.
Condition (i) is straight from the management knowledge about its own state, while (ii) can be based on empirical studies as in \cite{Carpaneto2008,Munkhammar2015} or in energy generation forecast based on weather as in \cite{BCDC}.

The system is said to be self-sufficient at time step $t$ if, and only if: $ S_\textup{min} - S(t-1)  < S(t) \leq S_\textup{max} - S(t-1)$.
The probability that this event happens is given in the following corollary.

\begin{corollary}
	The probability that the system is self-sufficient in time step $t$ is computed as: 
	$\mathrm{Pr}\left[ S_\textup{min} - S(t-1) < B(t) \leq  S_\textup{max} - S(t-1)  \right]$.	
\end{corollary}

In other words, the storage controller can compute the probabilities of these events only based on the distribution of $B(t) = G(t) - D(t)$, which is easily computed from the probability distribution functions \cite[pp. 185--186]{Papoulis}.

\section{Numerical examples}
\begin{figure}[t]
	\vspace{1ex}
	%
	\begin{subfigure}[b]{\columnwidth}
		\includegraphics[width=\columnwidth]{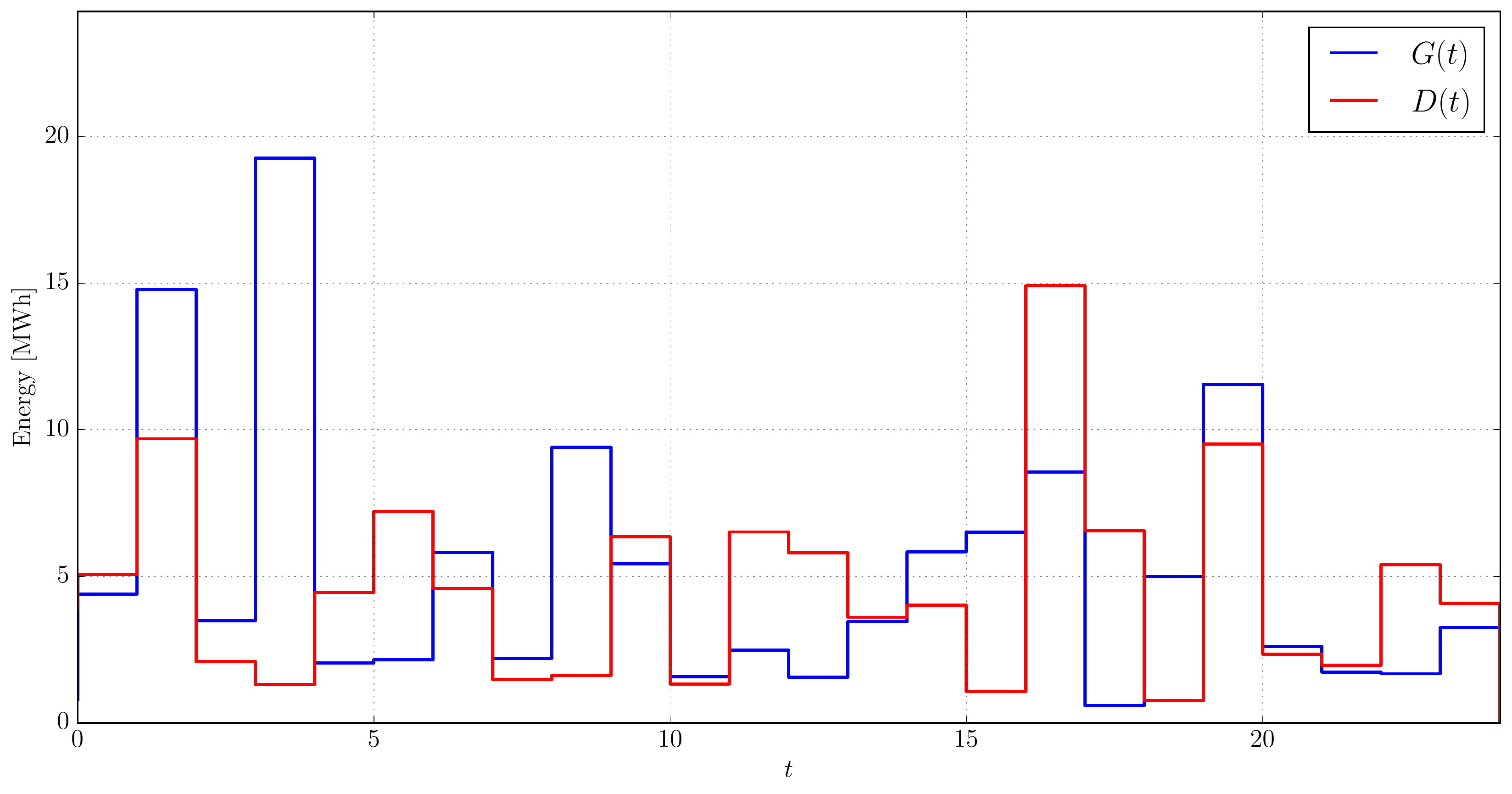}
		\caption{Realization of generation $G(t)$ and demand $D(t)$.}
		\label{fig:gen-dem}
		\vspace{1ex}
	\end{subfigure}
	\begin{subfigure}[b]{\columnwidth}
		\includegraphics[width=\columnwidth]{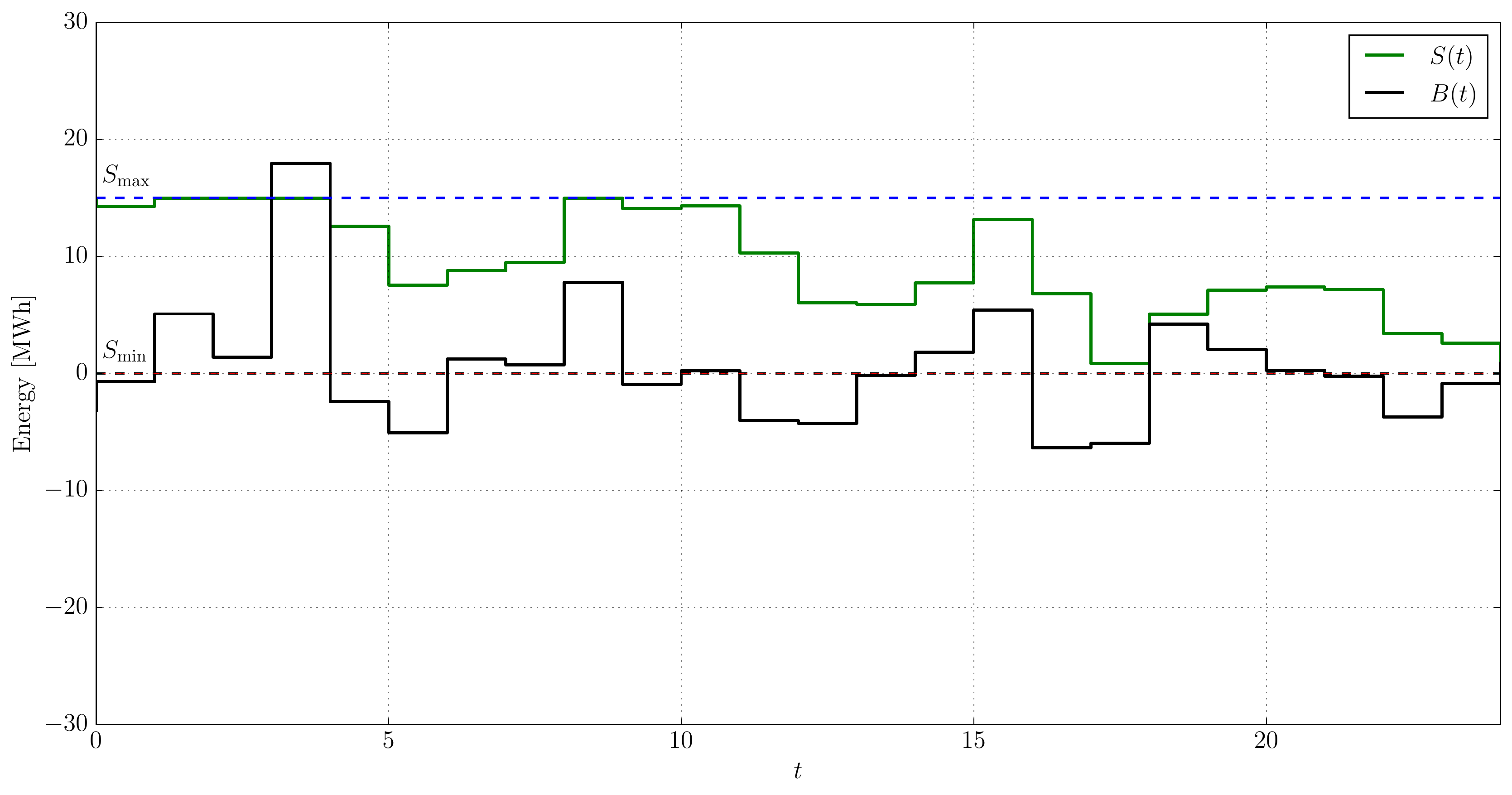}
		\caption{Balance function $B(t)$ and the storage situation $S(t)$.}
		\label{fig:storage}
	\end{subfigure}
	\caption{Example for $t=0,...,24$ (one day).The distribution probabilities of $G(t)$ and $D(t)$ are log-normal with arbitrary parameters. The code is available at \cite{github}.}
	\label{fig:example}
\end{figure}

We first assume the following system: a virtual utility that aggregates the demand of several households, and has under their use one wind farm as its generator and one hydro power plant as its storage.
We exemplify our scenario in Fig. \ref{fig:example} considering $G(t)$ and $D(t)$ as log-normal distribution with different parameters (refer to \cite{Papoulis})  for different time steps $t=0,...,24$.
The parameters where arbitrarily chosen with mean between $0.94 - 2.19$ MWh with variance ranging from $0.75$ to $1.125$ in the demand.
In generation the mean is $1.25$ MWh with variance $1$ regardless of $t$.

In the second scenario we consider a single household whose storage controlling entity needs to estimate the probability of not being self-sufficient during the next hour.
We assume that the household has a solar panel to generate electricity.
Assuming that the controller has a perfect weather forecast, the generation can be estimated with high accuracy \cite{BCDC}.
Therefore, $G(t+1)$ is not anymore a random variable.

As an example, we assume a battery with capacity $S_\textup{max} = 5$ kWh, and minimum level $S_\textup{min} = 0$.
The generation at $G(t+1) = G_{t+1} = 2$kWh (e.g. in a winter sunny day at 15:00)  is perfectly estimated by the weather forecast, while the the demand $D(t+1)$ is a Weibull distribution \cite{Munkhammar2015,Papoulis}:
\begin{equation}
\label{eq-weibull}
	\mathrm{Pr}\left[D(t+1) \leq y \right] =F_{D(t+1)}(y) = 1 - e^{-(y/\lambda)^k},
\end{equation}
where $\lambda < 0$ is the scale and $k < 0$ is the shape parameters.

The probability $p(S_t)$ of being self-sufficient in time $t+1$ as a function of the storage level $S(t) = S_t$ is given by \eqref{prop_storage_equation}.
Let $A$ denote the event when the generation plus storage will not be able to cover the demand and event $B$ when the generation surplus will overflow the storage capacity.
Then, $p(S_t) = 1 - p_A(S_t) - p_B(S_t)$.
From \eqref{eq-weibull}, we have:
\begin{align}
\vspace{-1ex}
\label{eq-weibull-numbers-A}
p_A(S_t) &=  1 - e^{-((G_{t+1} - S_t - S_\textup{min})/\lambda)^k}, \\
\label{eq-weibull-numbers-B}
p_B(S_t) &=  e^{-((G_{t+1} + S_t - S_\textup{max})/\lambda)^k}.
\end{align}
We present in Fig. \ref{fig:storage} a numerical example using  $\lambda = 2$ and $k=5$, given a mean of $\lambda\Gamma(1+1/k) \approx 2.1$ kWh.
\begin{figure}[t]
	\includegraphics[width=\columnwidth]{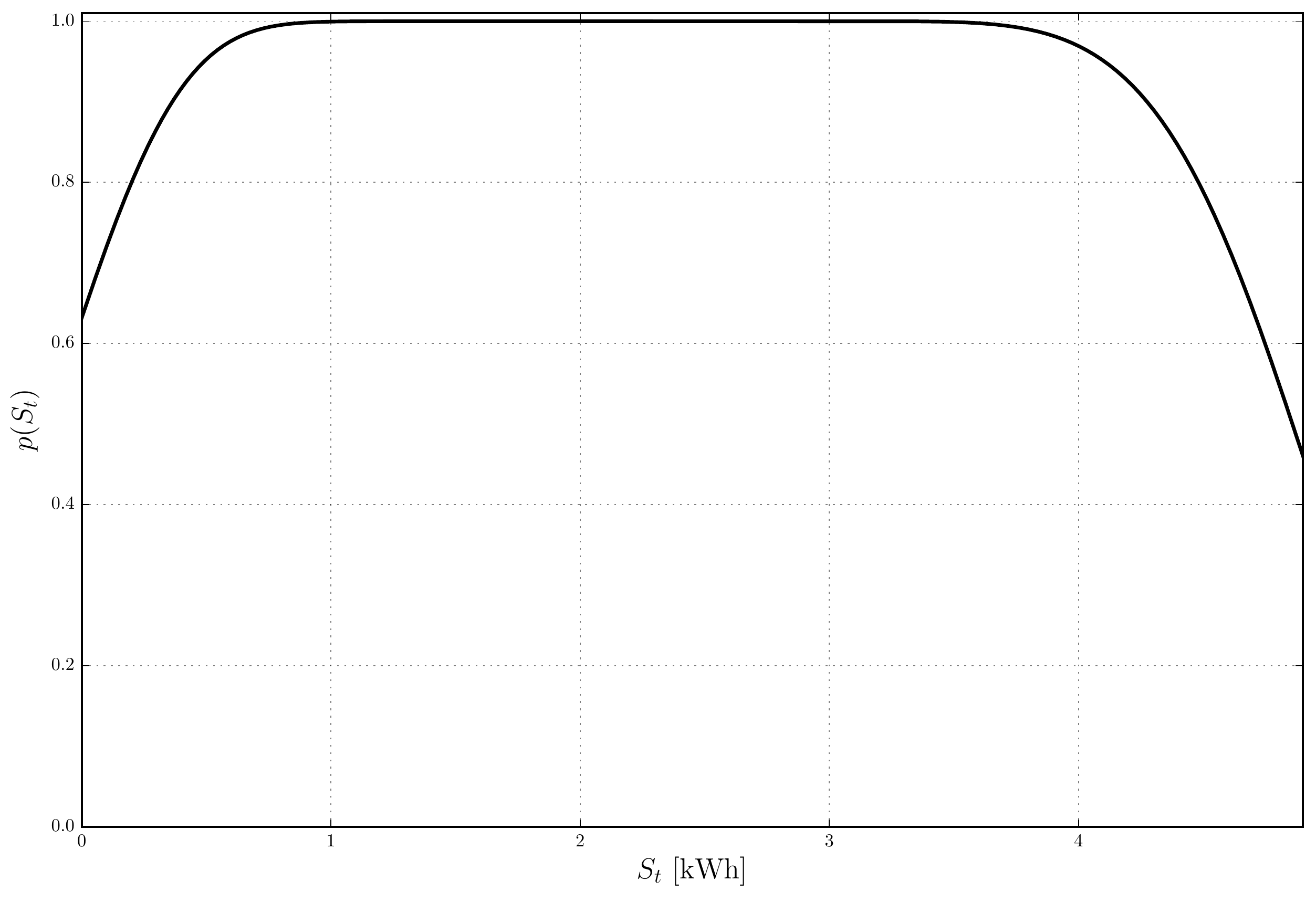}
	\caption{Probability $p(S_t)$ of not being self-sufficient as a function of the battery level $S(t)$, assuming that demand is a Weibull random variable.}
	\label{fig:storage}
	\vspace{-3ex}
\end{figure}

\vspace{-1ex}
\section{Discussions}
\vspace{-1ex}
This letter indicates how to evaluate whether a system composed by generation and storage can sustain the electricity demand for the next period.
For future research we plan to employ this approach as the basis for managing the situations where self-sufficiency is not reach.
We would like to check if a network of cooperatives can be a feasible option by sharing surpluses when deficits occur.

\vspace{-1ex}
\bibliographystyle{IEEEtran}

\end{document}